\expandafter\def\csname ver@fixltx2e.sty\endcsname{} 
\documentclass[letterpaper, 10 pt, conference]{ieeeconf}  

\IEEEoverridecommandlockouts                              
                                                          
\usepackage{cite}
\usepackage{float}
\usepackage{graphicx} 
\usepackage{tikz}
\usepackage{pgfplots}
\pgfplotsset{compat=1.15}
\usepackage{bm}
\usepackage{color}
\usepackage{xcolor}
\usepackage{amstext,epsfig,amssymb,amsbsy,amsmath}
\usepackage{mathtools}
\usepackage[output-decimal-marker={.}]{siunitx}
\usepackage{array}
\usepackage{tabularx}
\usepackage{multirow}
\usepackage{multicol}
\usepackage{setspace}
\let\labelindent\relax
\usepackage{enumitem}
\usepackage{dblfloatfix}
\usepackage{wrapfig}
\usepackage{url}
\usepackage{hyperref}
\usepackage{eurosym}
\usepackage{tablefootnote}
\usepackage{comment}
\usepackage{ifthen}
\usepackage{cases}
\newcommand{\mb}{\mathbf}
\usepackage[capitalise]{cleveref}
\usepackage{bm}
\usepgfplotslibrary{colorbrewer}

\expandafter\def\expandafter\UrlBreaks\expandafter{\UrlBreaks
  \do\a\do\b\do\c\do\d\do\e\do\f\do\g\do\h\do\i\do\j%
  \do\k\do\l\do\m\do\n\do\o\do\p\do\q\do\r\do\s\do\t%
  \do\u\do\v\do\w\do\x\do\y\do\z\do\A\do\B\do\C\do\D%
  \do\E\do\F\do\G\do\H\do\I\do\J\do\K\do\L\do\M\do\N%
  \do\O\do\P\do\Q\do\R\do\S\do\T\do\U\do\V\do\W\do\X%
  \do\Y\do\Z}

\usetikzlibrary{positioning,fit,arrows.meta,backgrounds}
\usetikzlibrary{patterns,shapes,positioning}
\usetikzlibrary{decorations.markings}
\tikzset{
    module/.style={%
        draw, rounded corners,
        minimum width=#1,
        minimum height=8mm,
        font=\sffamily
        },
    module/.default=2.1cm,
    >=LaTeX
}
\usetikzlibrary{calc}

\allowdisplaybreaks

\begin{document}

\bstctlcite{IEEEexample:BSTcontrol}

\title{Capturing Power System Dynamics by\\Physics-Informed Neural Networks and Optimization}%

\author{Georgios~S.~Misyris,~\IEEEmembership{Student~Member,~IEEE},  
        Jochen~Stiasny,~\IEEEmembership{Student~Member,~IEEE}, \\
        Spyros~Chatzivasileiadis,~\IEEEmembership{Senior Member,~IEEE}%
\thanks{G. S. Misyris, J. Stiasny and S. Chatzivasileiadis are with the Technical University of Denmark, Department of Electrical Engineering, Kgs. Lyngby, Denmark (emails: \{gmisy,jbest,spchatz\}@elektro.dtu.dk).}%
\thanks{This work is supported by Innovation Fund Denmark through the multiDC project (grant no. \mbox{6154-00020B}).}}

\maketitle

\vspace{-1em}
\begin{abstract}
This paper proposes a tractable framework to determine key characteristics of non-linear dynamic systems by converting physics-informed neural networks to a mixed integer linear program. Our focus is on power system applications. Traditional methods in power systems require the use of a large number of simulations and other heuristics to determine parameters such as the critical clearing time, i.e. the maximum allowable time within which a disturbance must be cleared before the system moves to instability. The work proposed in this paper uses physics-informed neural networks to capture the power system dynamic behavior and, through an exact transformation, converts them to a tractable optimization problem which can be used to determine critical system indices. 
By converting neural networks to mixed integer linear programs, our framework also allows to adjust the conservativeness of the neural network output with respect to the existing stability boundaries. We demonstrate the performance of our methods on the non-linear dynamics of converter-based generation in response to voltage disturbances.

\end{abstract}


\section{Introduction}
Modern power systems undergo a transition where large and centrally located synchronous generators are replaced by smaller and more distributed converter-based generation units. These changes substantially impact the dynamic behaviour of the power system and increase the number of phenomena that have to be taken into account for stability analyses \cite{Pico2019}. 
The particular case we investigate concerns `grid-following' converters, where the converter is controlled based on the measurement of the voltage of the external grid. The control action (i.e., setting an internal reference value) adapts to different control objectives and limitations, depending on the observed external voltage. 
A substantial voltage disturbance leads to a complex system response which affects the power output of the converter. If the disturbance becomes too large, the control scheme leads to a loss of the delivered power which cannot be recovered in the short-term. To avoid such undesirable system responses it is crucial to understand for which disturbance characteristics and control parameters such a substantial power loss occurs. 

Such an analysis of hybrid dynamical systems is notoriously difficult but unavoidable, as it is crucial to obtain insights on critical system responses. In power systems, these analyses often rely on a large number of time-domain simulations to exhibit the hybrid characteristic of the system. The high computational cost and the lack of providing informative margins towards the critical values are the main drawbacks. The former aspect can be partly addressed by the use of reduced-order modelling techniques \cite{Zali2013,Chaspierre2018,Chaspierre2020cigre} but these do not remove the second drawback. Direct methods using Lyapunov functions offer an alternative \cite{kundur1994power,pavella2012transient,Vu2015}, however, the results often lead to overly conservative conclusions and, except for \cite{Vu2015}, rely on system linearizations. A third route involves machine learning algorithms \cite{wehenkel2012automatic,donnot2017introducing, sun2018deep, arteaga2019deep, ferdin2019predicting} which can provide results at a fraction of the computational time. So far, their black-box nature and lack of interpretability of their assessments have been posing major barriers for adoption in a safety-critical environment. We aimed to remove these barriers with our recent work on neural network verification for power systems \cite{Venzke2020verification} and worst-case guarantees for neural network behaviour \cite{venzke2020learning}. Still, however, methods based on machine learning algorithms require the assessment of a large number of scenarios to obtain insights about critical system responses and estimate indices such as the critical clearing time or distance to instability.  

Main goal of this paper is to propose a method to determine critical indices for power systems (such as the critical clearing time) avoiding the need to perform thousands of simulations. To do that, we formulate an optimization problem that encodes in its constraints all the information of the underlying dynamical system. Instead of using linear approximations as in the direct methods \cite{kundur1994power,pavella2012transient}, we use Physics Informed Neural Networks (PINNs), which have shown good potential to capture power system dynamics governed by differential equations \cite{Misyris2019}. Inspired by our previous work \cite{Venzke2020verification},\cite{venzke2020learning}, we perform an exact transformation that converts the PINNs to a mixed integer linear program. This provides for a tractable optimization program that accurately captures in its constraints the power system dynamics, initially encoded in the PINN. Note that the reformulation of neural networks to an optimization program provides for a very flexible framework with a wide range of possible applications. Besides neural network verification \cite{tjeng2019evaluating, Venzke2020verification, venzke2020learning}, the same transformation has been recently used for the design of a neural network controller that ensures asymptotic stability \cite{karg2020stability}. 

The main contributions of this paper are: i) a rigorous framework that uses physics informed neural networks to capture the dynamics of power systems in a tractable optimization program, ii) the ability to determine critical indices of hybrid dynamic systems avoiding the need for exhaustive time-domain simulations, and iii) introducing the use of physics informed neural networks in such an optimization framework, which have the potential to capture power system dynamics more accurately than existing approximations.  

The proposed methodology in \cref{sec:methodology} splits the analysis task into two parts: 1) we use a PINN to approximate the dynamic system response; and 2) we reformulate the PINN approximation into a mixed-integer linear program which we subsequently use to rigorously investigate the system behaviour via optimization problems. We showcase this methodology in the context of the aforementioned grid-following converter dynamics that we introduce in detail in \cref{sec:case_study}.The results in \cref{sec:results} highlight the flexibility of the framework with respect to accommodating specific objectives of the system analysis. We conclude in \cref{sec:conclusion}.

\section{Methodology} \label{sec:methodology}
The problem at hand can be formulated as an optimization problem where we aim to find combinations of control inputs $\bm{u} \in \mathcal{U}$, system disturbances $\bm{w} \in \mathcal{W}$, and system parameters $ \bm{p} \in \mathcal{P}$ that yield the maximum value of an objective function $h(\bm{x}, \bm{y}, \bm{u}, \bm{w}, \bm{p})$ or simply a feasible point under a set of constraints. The constraints, representing element-wise lower $(\underline{\cdot})$ and upper bounds $(\overline{\cdot})$, can concern all differential states $\bm{x}$ and algebraic states $\bm{y}$ of the system as well as $ \bm{u}$, $\bm{w}$, and $\bm{p}$. Importantly, all variables are linked through a complex mapping $\phi(\bm{x}, \bm{y}, \bm{u}, \bm{w}, \bm{p})$. This leads to the optimization problem given by \eqref{eq:general_optimization}-\eqref{eq:complex_mapping}.

\begin{alignat}{2}
    \max_{\bm{u} \in \mathcal{U}, \bm{w} \in \mathcal{W}, \bm{p} \in \mathcal{P}} \quad &  h(\bm{x}, \bm{y}, \bm{u}, \bm{w}, \bm{p}) && \label{eq:general_optimization}   \\
    \text{s.t.}  \quad & \underline{\bm{u}} \leq  \bm{u} \leq \overline{\bm{u}}  && \\
    & \underline{\bm{x}} \leq \bm{x} \leq \overline{\bm{x}}  &&  \\
    & \underline{\bm{y}} \leq \bm{y} \leq \overline{\bm{y}}  &&  \\
    & \underline{\bm{u}} \leq \bm{u} \leq \overline{\bm{u}}  &&  \\
    & \underline{\bm{w}} \leq \bm{w} \leq \overline{\bm{w}}  &&  \\
    & \underline{\bm{p}} \leq \bm{p} \leq \overline{\bm{p}}  &&  \\
    & \phi(\bm{x}, \bm{y}, \bm{u}, \bm{w}, \bm{p}) \in \mathcal{S}\label{eq:complex_mapping} 
\end{alignat}

The mapping $\phi$ usually renders the optimization problem intractable, in particular if differential states and discrete events are involved. This is the case for hybrid dynamic systems as described by
\begin{align}
\frac{d}{dt}{\bm{x}} &= \bm{f}(\bm{x}, \bm{y}, \bm{u}, \bm{w}; \bm{p}) \\
 \bm{0} &= \bm{g}(\bm{x}, \bm{y}, \bm{u}, \bm{w}; \bm{p}) 
\end{align}
where $\bm{y}, \bm{u}, \bm{w}$ can show discrete switching behaviour.

The following methodology will describe how we can transform the potentially intractable mapping $\phi$ into a tractable approximation of it. This involves the approximation of the mapping by a PINN as described in \cref{subsec:NN_methodology} and, subsequently, its reformulation into a mixed-integer linear program in \cref{subsec:MILP_reformulation}.

\subsection{Function approximation via Physics-Informed Neural Networks}\label{subsec:NN_methodology}

The goal of the PINN is to accurately approximate the complex mapping $\phi$ such that we can obtain approximations of the system's state evolutions $\hat{\bm{x}}(t; \mathcal{C})$ and $\hat{\bm{y}}(t; \mathcal{C})$ given a set of characteristics $\mathcal{C}$.
\begin{align}
    \begin{bmatrix} \bm{x}(t; \mathcal{C})\\ \bm{y}(t; \mathcal{C})\end{bmatrix} \approx \begin{bmatrix} \hat{\bm{x}}(t; \mathcal{C})\\ \hat{\bm{y}}(t; \mathcal{C}) \end{bmatrix} = PINN(t; \mathcal{C})
\end{align}

These characteristics could be initial conditions $\bm{x}_0, \bm{y}_0$ but also control inputs, system disturbances, system parameters, all of which could be time-varying or constant. Note that the representation of $\mathcal{C}$ can be chosen freely as long as it yields a unique approximation $[\hat{\bm{x}}(t; \mathcal{C})^\top, \hat{\bm{y}}(t; \mathcal{C})^\top]^\top$. A constant control input $u$, for example, could be presented as a time-series $\mathcal{C} = [u(0), u(0.01), ...]^\top$ or the constant value itself $\mathcal{C} = [U]^\top$; it is a design choice that influences the training procedure and later on the formulation of the optimization problem.

\begin{figure}[ht]
    \centering
    \pagestyle{empty}
\def\layersep{1.4cm}
\def\nodeinlayersep{1cm}
\begin{tikzpicture}[
   shorten >=1pt,->,
   draw=black!70,
    node distance=\layersep,
    every pin edge/.style={<-,shorten <=1pt},
    neuron/.style={circle,fill=black!25,minimum size=17pt,inner sep=0pt},
    input neuron/.style={neuron, fill=blue!30, minimum size=20pt,draw=black},
    output neuron/.style={neuron, fill=red!50, minimum size=20pt,draw=black},
    hidden neuron/.style={neuron, fill=gray!30},
    operator neuron/.style={neuron, fill=yellow!50, minimum size=25pt,draw=black},
    summation neuron/.style={neuron, fill=gray!50, minimum size=16pt},
    parameter neuron/.style={neuron, fill=green!30, minimum size=25pt,draw=black},
    annot/.style={text width=4em, text centered}
]
    
    \node[input neuron] (I-1) at (0,-1.5*\nodeinlayersep) {t};  
    \node[input neuron] (I-2) at (0,-2.5*\nodeinlayersep) {$\mathcal{C}_1$};  
    \node (I-3) at (0,-3.5*\nodeinlayersep) {$\vdots$};  
    \node[input neuron] (I-4) at (0,-4.5*\nodeinlayersep) {$\mathcal{C}_l$};  

    \node[hidden neuron] (H1-1) at (1*\layersep,-1*\nodeinlayersep ) {$\sigma$};
    \node[hidden neuron] (H1-2) at (1*\layersep,-2*\nodeinlayersep ) {$\sigma$};
    \node[hidden neuron] (H1-3) at (1*\layersep,-3*\nodeinlayersep ) {$\sigma$};
    \node (H1-4) at (1*\layersep,-4*\nodeinlayersep ) {$\vdots$};
    \node[hidden neuron] (H1-5) at (1*\layersep,-5*\nodeinlayersep ) {$\sigma$};
    
    \node[hidden neuron] (H2-1) at (2*\layersep,-1*\nodeinlayersep ) {$\sigma$};
    \node[hidden neuron] (H2-2) at (2*\layersep,-2*\nodeinlayersep ) {$\sigma$};
    \node[hidden neuron] (H2-3) at (2*\layersep,-3*\nodeinlayersep ) {$\sigma$};
    \node (H2-4) at (2*\layersep,-4*\nodeinlayersep ) {$\vdots$};
    \node[hidden neuron] (H2-5) at (2*\layersep,-5*\nodeinlayersep ) {$\sigma$};
    
    \node[output neuron] (O-1) at (3*\layersep,-1*\nodeinlayersep) {$\hat{x}^1$};
    \node (O-2) at (3*\layersep,-1.6*\nodeinlayersep) {$\vdots$};
    \node[output neuron] (O-3) at (3*\layersep,-2.4*\nodeinlayersep) {$\hat{x}^m$};
    \node[output neuron] (O-4) at (3*\layersep,-3.5*\nodeinlayersep) {$\hat{y}^1$};
    \node (O-5) at (3*\layersep,-4.1*\nodeinlayersep) {$\vdots$};
    \node[output neuron] (O-6) at (3*\layersep,-4.9*\nodeinlayersep) {$\hat{y}^n$};

    \node[operator neuron] (D-1) at (4.2*\layersep,-1.5*\nodeinlayersep) {$\hat{\bm{x}}$};
    \node[operator neuron] (D-2) at (4.2*\layersep,-3*\nodeinlayersep) {$\frac{\partial}{\partial t} \hat{\bm{x}}$};
    \node[operator neuron] (D-3) at (4.2*\layersep,-4.5*\nodeinlayersep) {$\hat{\bm{y}}$};

   
    \node [rounded corners=0.2cm] (loss_x) at (5.3*\layersep,-1.1*\nodeinlayersep) [draw,fill=orange!30,thick,minimum width=0.8cm,minimum height=0.8cm] {$\sum \mathcal{L}_x^i$};
     \node [rounded corners=0.2cm] (loss_f) at (5.3*\layersep,-2.4*\nodeinlayersep) [draw,fill=orange!30,thick,minimum width=0.8cm,minimum height=0.8cm] {$\sum \mathcal{L}_f^i$};
    \node [rounded corners=0.2cm] (loss_g) at (5.3*\layersep,-3.6*\nodeinlayersep) [draw,fill=orange!30,thick,minimum width=0.8cm,minimum height=0.8cm] {$\sum \mathcal{L}_g^i$};
   \node [rounded corners=0.2cm] (loss_y) at (5.3*\layersep,-4.9*\nodeinlayersep) [draw,fill=orange!30,thick,minimum width=0.8cm,minimum height=0.8cm] {$\sum \mathcal{L}_y^i$};
   
   

    \foreach \source in {1, 2, 4}
        \foreach \dest in {1,...,3,5} 
            \path (I-\source) edge (H1-\dest);
    
    \foreach \source in {1,...,3,5}
        \foreach \dest in {1,...,3,5} 
            \path (H1-\source) edge (H2-\dest);
            
    \foreach \source in {1,...,3,5}
        \foreach \dest in {1,3,4,6} 
            \path (H2-\source) edge (O-\dest);
            
    \foreach \source in {1,3} 
        \foreach \dest in {1,2}
            \path (O-\source) edge (D-\dest);

    \foreach \source in {4,6} 
        \foreach \dest in {3}
            \path (O-\source) edge (D-\dest);
    
    \path (D-1) edge (loss_x); 
    \path (D-1) edge (loss_f); 
    \path (D-1) edge (loss_g);
    \path (D-2) edge (loss_f);
    \path (D-3) edge (loss_f);
    \path (D-3) edge (loss_g); 
    \path (D-3) edge (loss_y); 
    
    
    \node [rounded corners=0.2cm] (NeuralNetwork) at (1.5*\layersep,-3*\nodeinlayersep ) [draw,black, dashed, very thick,minimum width=4*\layersep,minimum height=5*\nodeinlayersep] {};
    \node [rounded corners=0.2cm] (AD) at (4.2*\layersep,-3*\nodeinlayersep) [draw,black, dashed, very thick,minimum width=1*\layersep,minimum height=5*\nodeinlayersep] {};
    \node[annot, above=2pt] at (NeuralNetwork.north) {NN};
    \node[annot, above=2pt] at (AD.north) {AD};
\end{tikzpicture} 
    \caption{Physics-informed neural network architecture (adapted from \cite{stiasny2020physicsinformed}), consisting of a dense neural network (NN), followed by applying automatic differentiation (AD) to the differential states $\hat{x}$. The calculation of the losses is based on data ($\mathcal{L}_x^i, \mathcal{L}_y^i$) and on the governing equations ($\mathcal{L}_f^i, \mathcal{L}_g^i$).}
    \label{fig:pinn_architecture}
\end{figure}
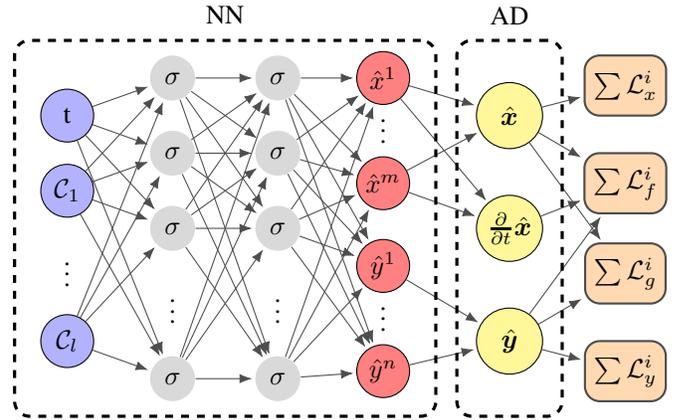

The PINN we use consists of a multi-layer perceptron network with the following structure, which is also illustrated in \cref{fig:pinn_architecture}:

\begin{alignat}{2}
    [t, \mathcal{C}]^\top &= \bm{z}_0 &&\label{eq:NN_input}\\
    \hat{\bm{z}}_{k+1} &= \bm{W}_{k+1} \bm{z}_k + \bm{b}_{k+1} && \forall k = 0, 1, ..., K-1\label{eq:NN_hidden_layers}\\
    \bm{z}_k &= \max (\hat{\bm{z}}_k, 0) && \forall k = 1, ..., K\label{eq:ReLUactivation} \\
    [\hat{\bm{x}}^\top, \hat{\bm{y}}^\top]^\top &= \bm{W}_{K+1} \bm{z}_K + \bm{b}_{K+1}&&\label{eq:NN_output}.
\end{alignat}
The choice of the $\max$-operator in \eqref{eq:ReLUactivation} - also known as Rectified Linear Unit (ReLU) - as the non-linear activation function is essential for the reformulation of the PINN in \cref{subsec:MILP_reformulation}. Other common activation functions such as the hyperbolic tangent do not allow this step. $\bm{W}_i$ and $\bm{b}_i$ represent the weight matrices and bias vectors in each hidden layer and they are the adjustable parameters during the training. In order to determine $\bm{W}_i$ and $\bm{b}_i$ we minimize the mismatch between the approximated states $\hat{\bm{x}}(t; \mathcal{C})$, $\hat{\bm{y}}(t; \mathcal{C})$ and a dataset that contains simulated values $\bm{x}(t; \mathcal{C})$, $\bm{y}(t; \mathcal{C})$ which form the ground truth. The mean-squared error across $N$ data points from the dataset yield a loss term for each of the $m$ differential and $n$ algebraic states 
\begin{alignat}{2}
    \mathcal{L}_{x}^i &= \frac{1}{N} \sum_{j=1}^N (x_j^i - \hat{x}_j^i)^2 \qquad && \forall i = 1, ..., m\\
    \mathcal{L}_{y}^i &= \frac{1}{N} \sum_{j=1}^N (y_j^i - \hat{y}_j^i)^2 && \forall i = 1, ..., n.
\end{alignat}
On top of these loss terms based on a dataset we add the physical knowledge of the system in form of a regularisation - this addition differentiates a PINN from a plain NN and allows for a more data efficient process. We use the regularisation to evaluate whether the state approximations satisfy the governing differential and algebraic equations. The associated loss terms $\mathcal{L}_f^i, \mathcal{L}_g^i$ take the form
\begin{alignat}{2}
    \mathcal{L}_{f}^i &= \frac{1}{N_c} \sum_{j=1}^{N_c} (f(\bm{\hat{x}}_j, \bm{\hat{y}}_j, \bm{u}_j, \bm{w}_j; \bm{p})^i - \frac{d}{dt}\hat{x}_j^i)^2 \; && \forall i = 1, ..., m\\
    \mathcal{L}_{g}^i &= \frac{1}{N_c} \sum_{j=1}^{N_c} (g(\bm{\hat{x}}_j, \bm{\hat{y}}_j, \bm{u}_j, \bm{w}_j; \bm{p})^i)^2 && \forall i = 1, ..., n.
\end{alignat}
Note that to evaluate these loss terms we do not require to know $\bm{x}$ and $\bm{y}$, hence, we can evaluate $\mathcal{L}_{f}^i$ and $\mathcal{L}_{g}^i$ at $N_c$ collocation points across the input domain without the need for additional simulations. We only require the approximations $\hat{\bm{x}}, \hat{\bm{y}}$ and the term $\frac{d}{dt}\hat{x}_j^i$ can be computed by applying automatic differentiation \cite{baydin2015automatic} on the differential states $\hat{\bm{x}}$ with respect to the input variable $t$. Lastly, the values of $\bm{u}_j, \bm{w}_j, \bm{p}$ in the function evaluations $f$ and $g$ are either given as fixed parameters or can be inferred from the input characteristics $\mathcal{C}$.

The weighted sum\footnote{For the values of the weighing parameters $\lambda_x^i, \lambda_y^i, \lambda_f^i, \lambda_g^i$ please refer to the published code on \url{https://github.com/gmisy}} of the losses then forms the objective function for the training procedure

\begin{align}
    \min_{\bm{W}_i, \bm{b}_i} \quad &\lambda_x^i \mathcal{L}_{x}^i + \lambda_y^i \mathcal{L}_{y}^i + \lambda_f^i \mathcal{L}_{f}^i + \lambda_g^i \mathcal{L}_{g}^i\\
    \text{s.t.} \quad & \eqref{eq:NN_input} - \eqref{eq:NN_output}.
\end{align}

\subsection{Mixed-Integer Linear reformulation}\label{subsec:MILP_reformulation}

The above described PINN is effectively a combination of linear transformations and non-linear activation functions, here ReLUs \eqref{eq:ReLUactivation}. By using the exact reformulation \eqref{Eq:Milp1}-\eqref{Eq:Milp5} of the ReLUs \cite{tjeng2019evaluating} we convert this general non-linear program into a mixed-integer linear program (MILP) which in turns allows us to benefit from the specialized MILP solvers. 


\begin{subnumcases}{\bm{z}_k = \max(\hat{\bm{z}}_k,0)\Rightarrow}
\bm{z}_k  \leq \hat{\bm{z}}_k - \hat{\bm{z}}^{\text{min}}_k  (1-\mb{b}_k) \label{Eq:Milp1} \\ 
\bm{z}_k  \geq \hat{\bm{z}}_k \label{Eq:Milp2}   \\
\bm{z}_k  \leq \hat{\bm{z}}^{\text{max}}_k \mb{b}_k  \label{Eq:Milp3}  \\
\bm{z}_k   \geq \bm{0}  \label{Eq:Milp4}  \\
\mb{b}_k \in \{0,1\}^{N_k} \label{Eq:Milp5}
\end{subnumcases}

The reformulation hinges around the use of a binary variable $\mb{b}_k$ for each ReLU, i.e. for the $N_k$ neurons in each layer of the PINN, and works as follows. If $\mb{b}_k = 0$, \eqref{Eq:Milp3} and \eqref{Eq:Milp4} demand $\bm{z}_k = 0$, and if $\mb{b}_k = 1$, \eqref{Eq:Milp1} and \eqref{Eq:Milp2} yield $\bm{z}_k = \hat{\bm{z}}_k$. $\hat{\bm{z}}^{\text{min}}_k$ and $\hat{\bm{z}}^{\text{max}}_k$ are required to bound the problem, however, they need to be chosen large enough to not be binding. Tightening these bounds, by means of interval arithmetic \cite{tjeng2019evaluating} and a bound tightening algorithm \cite{venzke2020learning}, can reduce the computation time significantly.

\section{Case study} \label{sec:case_study}
This section introduces the essential aspects of the converter dynamics which we seek to analyse; an extensive description can be found in \cite{Chaspierre2020, Chaspierre2020cigre}. Furthermore, we describe the training of the PINNs as well as the formulation of the optimization problems through which we analyze the system characteristics in response to a voltage disturbance. Our objective is to determine: i) the maximum allowable disturbance duration $\Delta T$, given the voltage disturbance magnitude $\Delta V$, without entering an undesired control scheme, and ii) the amount of active power that is delivered shortly after the disturbance is cleared. 

\subsection{Current control in grid-following converters}\label{subsec:converter_dynamics}

\begin{figure}[t]
    \centering
    \pagestyle{empty}
\def\subplotWidth{4.5cm}
\def\horizontalDistance{6cm}
\def\subplotHeight{5.5cm}
\def\heightOffset{-3.5cm}

    \begin{tikzpicture}[     every node/.style={font=\footnotesize}]

    \begin{axis}[
    xlabel near ticks, 
    ylabel near ticks, 
    width=\subplotWidth, 
    height=\subplotHeight,
    yshift=0,
    xshift=0,
    line width=0.5,
    xlabel={$V_{\rm meas}$},
    ymax = 1.2,
    ymin= 0.0,
    xmin=0.0, xmax=1.2,
    axis x line=bottom,
    axis y line=left,
    ylabel={$i_Q$},
    ytick ={0.2, 0.9},
    yticklabels={%
        $I_{Q0}$,%
         $I_{\rm nom}$,%
    },
    xtick={0.9},
    xticklabels={%
         $V_Q$,%
    },
    xlabel style = {at={(axis description cs:1,0)},anchor=north},
    ylabel style = {at={(axis description cs:0,1)},anchor=east, rotate=-90},
    thick,
    %
    ]
    
    \draw[black] (0, 0.9) -- (0.2, 0.9) -- (0.9, 0.2);
    \draw[black, dashed] (0.0, 0.2) -- (0.9, 0.2) -- (0.9, 0);
    \draw[black, dashed] (0.5, 0.6) -- (0.85, 0.6);
    
    \draw [black, dashed, thin] (0.75, 0.6) arc (0:-45:0.25);
    \node at (0.9, 0.5) {$K_{RCI}$};
    \end{axis}
    
        \begin{axis}[
    xlabel near ticks, 
    ylabel near ticks, 
    width=\subplotWidth, 
    height=3.5cm,
    xshift=3.5cm,
    yshift=2cm,
    line width=0.5,
    xlabel={$V_{\rm meas}$},
    ymax = 1.2,
    ymin= 0.0,
    xmin=0.0, xmax=1.1,
    axis x line=bottom,
    axis y line=left,
    ylabel={$f$},
    ytick ={0, 0.3, 0.6, 1},
    yticklabels={0, $f_{\rm post}$, $c$, 1},
    xtick={0.3, 0.7},
    xticklabels={%
         $V_{\rm min}$,%
          $V_{\rm int}$,%
    },
    xlabel style = {at={(axis description cs:1,0)},anchor=west},
    ylabel style = {at={(axis description cs:0,1)},anchor=east, rotate=-90},
    thick,
    %
    ]
    
    \draw[black, dashed] (0,1) -- (0.7,1);
    \draw[black, dashed] (0,0.6) -- (0.7,0.6);
    \draw[red, dashed] (0,0.3) -- (0.5,0.3) -- (0.5, -1.0);
    \begin{scope}[decoration={
    markings,
    mark=at position 0.7 with {\arrow{>}}}
    ] 
    \draw[postaction={decorate}] (1,1)--(0.7,1);
    \draw[postaction={decorate}] (0.7,1)--(0.7,0.6);
    \end{scope}
    \begin{scope}[decoration={
    markings,
    mark=at position 0.4 with {\arrow{>}}}
    ] 
    \draw[postaction={decorate}] (0.7,0.6)--(0.3, 0.0);
    \end{scope}
    \end{axis}
    
            \begin{axis}[
    xlabel near ticks, 
    ylabel near ticks, 
    width=\subplotWidth, 
    height=3cm,
    xshift=3.5cm,
    yshift=0cm,
    line width=0.5,
    xlabel={$V_{\rm meas}$},
    ymax = 0.0,
    ymin= -0.8,
    xmin=0.0, xmax=1.1,
    axis x line=top,
    axis y line=left,
    ylabel={$t$},
    xmajorticks=false,
    ymajorticks=false,
    xlabel style = {at={(axis description cs:1,1)},anchor=west},
    ylabel style = {at={(axis description cs:0,0)},anchor=south east, rotate=-90},
    thick,
    y axis line style={stealth-},   
               ]
    %
    ]

    \draw[black] (1, 0) -- (1, -0.2);  
    \draw [black] plot [smooth, tension=0.8] coordinates { (1,-0.2) (0.9,-0.21) (0.5,-0.3) (0.85, -0.45) (0.95,-0.6) (1.0,-0.8)};
    \draw[red, dashed] (0.5,0.0) -- (0.5, -0.3);

    \end{axis}
    \end{tikzpicture}
    \caption{Adjustment of $i_Q$ and $f$ based on the voltage level. $i_Q$ models the voltage support characteristic and $f$ the Low Voltage Ride-Through (LVRT) characteristic. Adapted from \cite{Chaspierre2020cigre}.}
    \label{fig:CurrentController}
\end{figure}
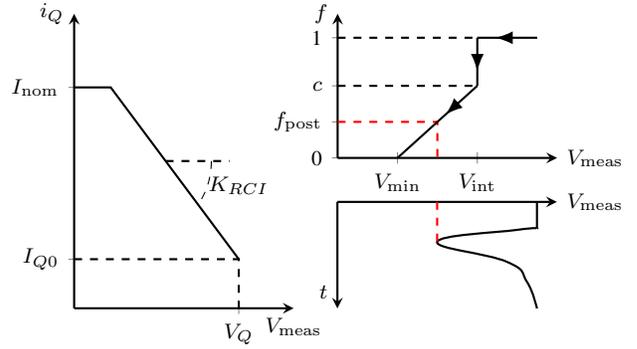

The following model stems from the model-reduction approach laid out in \cite{Chaspierre2020}, however, here we shall only focus on the control of the converter current for brevity. The differential states of the currents $i_d, i_q$ lag their respective reference values $i_{d}^{\rm ref}, i_{q}^{\rm ref}$.
\begin{align}
    \frac{d i_{d}}{dt} &= \frac{1}{T_p}(i_{d}^{\rm ref} - i_{d}), \\
    \frac{d i_{q}}{dt} &= \frac{1}{T_q}(i_{q}^{\rm ref} - i_{q}).
\end{align}
These reference values are defined by the desired active and reactive power output $P_{{\rm ext}}, Q_{{\rm ext}}$ under normal operating conditions
\begin{align}
    i_{d}^{\rm ref} & = \left(\frac{P_{{\rm ext}}}{V_{\rm meas}}\right) f, \\
    i_{q}^{\rm ref} & = \left( -\frac{Q_{{\rm ext}}}{V_{\rm meas}} + i_Q \right) f,
\end{align}
and they are adjusted by a current injection $i_Q$ for voltage support
\begin{equation}\label{eq:voltage_support}
i_{Q}= \begin{cases}
0, & V_{\rm meas} > V_Q\\
- \left(K_{RCI}(V_Q-V_{\rm meas}) + I_{Q0}\right), & V_{\rm meas} \leq V_Q
\end{cases}
\end{equation}
and the factor $f$ to represent the Low Voltage Ride-Through (LVRT) capabilities in case of abnormal conditions
\begin{equation}\label{eq:partial_tripping}
f= \begin{cases}
1, & V_{\rm meas} \geq V_{\rm int} \\
c \; \frac{V_{\rm meas} - V_{\rm min}}{V_{\rm int} - V_{\rm min}}, & V_{\rm min} \leq V_{\rm meas} < V_{\rm int},\\
0, & V_{\rm meas} < V_{\rm min}.
\end{cases}
\end{equation}
\Cref{fig:CurrentController} illustrates the control mechanism and shows how the LVRT characteristic leads to a the reduction of $f$ to $f_{\rm post}$ which is irreversible in the short term. A further complication arises from the current limiters:
\begin{equation}\label{eq:current_limit_d}
I_{d}^{\rm max}= \begin{cases}
I_{\rm nom}, & V_{\rm meas} \geq 0.9\\
\sqrt{I_{\rm nom}^2 - i_q^2}, & V_{\rm meas} < 0.9
\end{cases}
\end{equation}

\begin{equation}\label{eq:current_limit_q}
I_{q}^{\rm max}= \begin{cases}
\sqrt{I_{\rm nom}^2 - i_d^2}, & V_{\rm meas} \geq 0.9\\
I_{\rm nom}, & V_{\rm meas}  < 0.9
\end{cases}
\end{equation}

These control actions all depend on the measured voltage $V_{\rm meas}$ which constitutes another differential state, and which is linked to the external voltage through a low-pass filter in the transformed voltages $v_d, v_q$:

\begin{equation}
    \frac{d V_{\rm meas}}{dt} = \frac{1}{T_m}(\sqrt{v_d^2 + v_q^2} - V_{\rm meas})
\end{equation}
The delivered active and reactive power then depend on the external voltages and the converter currents. 
\begin{align}
    P_{\rm total} &= v_{g,d}i_{d} + v_{g,q}i_{q}, \\
    Q_{\rm total} &= v_{g,q}i_{d} - v_{g,d}i_{q}. \label{Eq:Q}
\end{align}
Appendix A contains the remaining system equations. 

\subsection{PINN training}
The characteristics $\mathcal{C}$ we use in order to encode the response to a voltage disturbance are the voltage disturbance duration $\Delta T$ and its magnitude $\Delta V$. The input $\bm{z}_0 = [t, \Delta V, \Delta T]^\top$ is followed by three hidden layers with $N_k = 50$ neurons per layer. The output is given by $[\hat{\bm{x}}^\top, \hat{\bm{y}}^\top]^\top$ with
\begin{equation}
    \hat{\bm{x}} = \begin{bmatrix}
    \hat{\theta}_{{\rm pll}} & \hat{i}_{d} & \hat{i}_{q} & \hat{V}_{\rm meas}
    \end{bmatrix}^\top \label{Eq:x_vector}
\end{equation}
{\footnotesize
\begin{align}
    \hat{\bm{y}} = 
    \Big[\hat{v}_{d}, \> \hat{v}_{q},\> \hat{\omega}_{{\rm pll}},\> \hat{P}_{\rm VSC}, \>  \hat{V}_{\rm PCC}, \> \hat{Q}_{\rm VSC}, \> \hat{v}_{g,d}, \> \hat{v}_{g,q}  \> P_{\rm total}, \> Q_{\rm total}\Big]^\top  \label{Eq:y_vector}.
\end{align}
}

The PINN is implemented using Tensorflow \cite{abadi2016tensorflow} and is trained on a dataset that consists of trajectories with a time step size of 0.001s up to 1s. 40 trajectories are simulated with 
\begin{align}
    \Delta T \in &\{0.1, 0.15, 0.2, 0.25\}s\\
    \Delta V \in &\{0.2, 0.267, 0.333, 0.4, 0.467, \\
    &0.533, 0.6, 0.667, 0.733, 0.8\} pu
\end{align}
with the initial condition being the equilibrium for the converter set-point 
\begin{align}
    \bm{u} = [P_{{\rm ext}}, Q_{{\rm ext}}] = [0.8, 0.2] pu.
\end{align}
In order to evaluate the governing equations at the collocation points we require the values of the external voltage $\bm{w} = [V_t]$. The time-series for $V_t$ is constructed by subtracting $\Delta V$ for a duration of $\Delta T$ from the set voltage $V_t = 1$pu. The system parameters $\bm{p}$ are not altered in this case study. All computations were performed on a regular machine (i5-7200U CPU @ 2.50GHz, 16GB RAM).

\subsection{Optimization problem formulation}\label{subsec:optimization_problems}
We consider the problem of finding the maximum allowable disturbance duration while not entering the low-voltage ride through (LVRT) mode, i.e. $V_{\rm meas} < V_{\rm int}$, since the delivered power can return to the desired power set-point after the disturbance is cleared. Otherwise, the factor $f$ leads to lower reference currents $i_{d}^{\rm ref}, i_{q}^{\rm ref}$ and hence to a reduced delivered power in the short-term. The presented framework provides the flexibility to analyze this question from multiple angles.

First, we require that \eqref{eq:min_voltage} is satisfied, i.e. the LVRT mode is never entered. We add a parameter $\epsilon$ to control how close the approximation can be to the critical voltage $V_{\rm int}$. By adding this buffer, we account for the fact that the PINN yields only an approximation of the system dynamics. This effectively shrinks the feasible space, which can also be seen as controlling the conservativeness of the resulting stability assessment. The resulting optimization problem reads as
\begin{alignat}{2}
\max_{\begin{smallmatrix}t,\hat{\bm{x}}, \hat{\bm{y}}, \bm{z}, \mb{b}\end{smallmatrix}} \, \, & \Delta T && \label{eq:obj_maximumDT_1}\\
    \text{s.t.}  \quad & V_{\rm int} + \epsilon \leq \hat{V}_{\rm meas} \label{eq:min_voltage} \\
    & t = \Delta T \label{eq:T_DT}\\
    & \Delta {V} = \Delta {V}' \label{eq:V_DV} \\
    & \eqref{eq:NN_input}, \eqref{eq:NN_hidden_layers},\eqref{eq:NN_output}\\
    & \eqref{Eq:Milp1}-\eqref{Eq:Milp5}. \label{eq:obj_maximumDT_2}
\end{alignat}    
and then we solve it for different magnitudes $\Delta {V}'$ of the disturbance. Constraint \eqref{eq:T_DT} is added as the lowest voltage will occur right before the voltage disturbance is cleared.

For the second perspective on the problem, we consider the final active power delivery $\hat{P}_{total}$ after 1s. By controlling $\mu \in [0;1]$ in \eqref{eq:mu_P_ext} we can analyze the power loss after $t=1$s for a given $\Delta {V}'$
\begin{alignat}{2}
\max_{\begin{smallmatrix}t,\hat{\bm{x}},\hat{\bm{y}}, \bm{z}, \mb{b}\end{smallmatrix}} \, \, & \Delta T && \label{eq:obj_maximumDT_3}\\
    \text{s.t.}  \quad & \hat{P}_{total} \geq \mu \, P_{{\rm ext}} \label{eq:mu_P_ext}\\
    & t = 1s \label{eq:T_DT_2}\\
    & \Delta {V} = \Delta {V}' \label{eq:V_DV_2} \\
    & \eqref{eq:NN_input}, \eqref{eq:NN_hidden_layers},\eqref{eq:NN_output}\\
    & \eqref{Eq:Milp1}-\eqref{Eq:Milp5} \label{eq:obj_maximumDT_4}
\end{alignat}    
The solution of the optimization problems is performed in Matlab \cite{MATLAB:R2019b_u2} using YALMIP \cite{Lofberg2004} and a MILP solver provided by Gurobi \cite{gurobi}.

\section{Results} \label{sec:results}
In this section we present the resulting approximations of the PINN and the analysis of the dynamic system based on the optimization problems in \cref{subsec:optimization_problems}.

\subsection{PINN approximation}
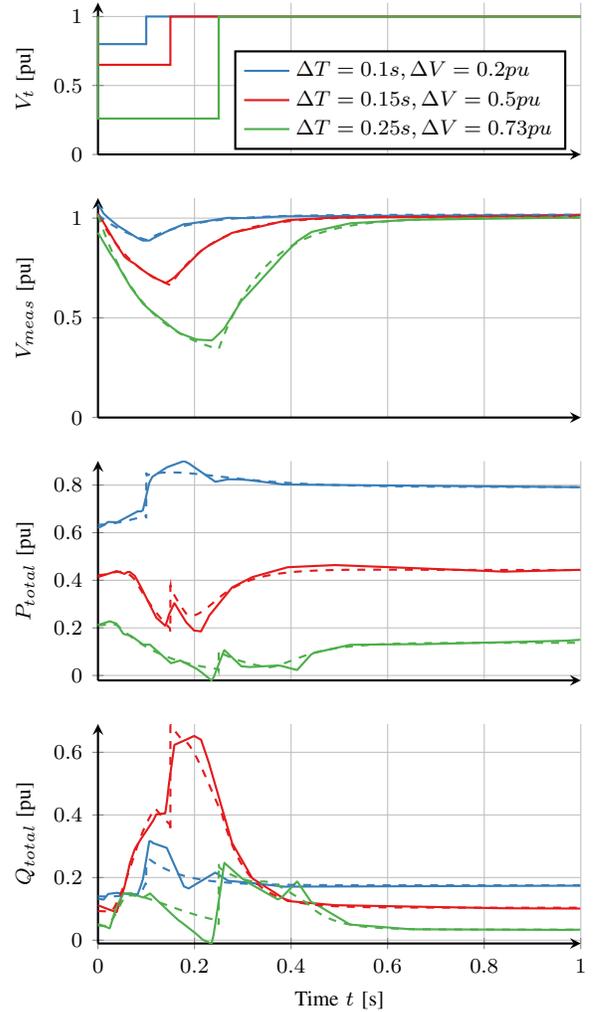
\begin{figure}[t]
    \centering
    \pagestyle{empty}
\def\subplotWidth{8cm}
\def\horizontalDistance{5cm}
\def\subplotHeight{4.5cm}
\def\heightOffset{-3.5cm}

\definecolor{blue_characteristic}{HTML}{377EB8}
\definecolor{red_characteristic}{HTML}{E41A1C}
\definecolor{green_characteristic}{HTML}{4DAF4A}
    \begin{tikzpicture}[     every node/.style={font=\footnotesize}]

    \begin{axis}[
    height=0.8*\subplotHeight,
    width=\subplotWidth,
    yshift=-\heightOffset,
    xmajorgrids=true,
    ymajorgrids=true,
    ymin=0,ymax=1.1,
    xmin=0, xmax=1,
    axis y line=left,
    x axis line style=-,
    ylabel={$V_{t}$ [pu]},
    grid=major, 
    axis x line=bottom,
    xmajorticks=false,
    thick,
    legend pos= south east,
    legend cell align={left}
    ]

    \addplot[draw=blue_characteristic] coordinates {(0,1) (0, 0.8) (0.1, 0.8) (0.1, 1.0) (1,1)};
    \addplot[draw=red_characteristic] coordinates { (0,1) (0, 0.65) (0.15, 0.65) (0.15, 1.0) (1,1)};
    \addplot[draw=green_characteristic] coordinates { (0,1) (0, 0.26) (0.25, 0.26) (0.25, 1.0) (1,1)};
     
    \addlegendentry{$\Delta T = 0.1s, \Delta V = 0.2pu$}
    \addlegendentry{$\Delta T = 0.15s, \Delta V = 0.5pu$}
    \addlegendentry{$\Delta T = 0.25s, \Delta V = 0.73pu$}
    \end{axis}

    \begin{axis}[
    height=\subplotHeight,
    width=\subplotWidth,
    xmajorgrids=true,
    ymajorgrids=true,
    ymin=0,ymax=1.1,
    xmin=0, xmax=1,
    axis y line=left,
    x axis line style=-,
    ylabel={$V_{meas}$ [pu]},
    grid=major, 
    axis x line=bottom,
    xmajorticks=false,
    smooth,
    thick,
    ]

    \addplot[draw=blue_characteristic] table[x index=0, y index=5] {data/trajectory_0_1s_0_2_pu.dat};
    \addplot[draw=red_characteristic] table[x index=0, y index=5] {data/trajectory_0_15s_0_5_pu.dat};
    \addplot[draw=green_characteristic] table[x index=0, y index=5] {data/trajectory_0_25s_0_73_pu.dat};

    \addplot[draw=blue_characteristic, dashed] table[x index=0, y index=22] {data/trajectory_0_1s_0_2_pu.dat};
    \addplot[draw=red_characteristic, dashed] table[x index=0, y index=22] {data/trajectory_0_15s_0_5_pu.dat};
    \addplot[draw=green_characteristic, dashed] table[x index=0, y index=22] {data/trajectory_0_25s_0_73_pu.dat};
    \end{axis}
    
    \begin{axis}[
    height=\subplotHeight,
    width=\subplotWidth,
    yshift=\heightOffset,
    xmajorgrids=true,
    ymajorgrids=true,
    xmin=0, xmax=1,
    axis y line=left,
    x axis line style=-,
    ylabel={$P_{total}$ [pu]},
    grid=major, 
    axis x line=bottom,
    xmajorticks=false,
    smooth,
    thick,
    ]

    \addplot[draw=blue_characteristic] table[x index=0, y index=16] {data/trajectory_0_1s_0_2_pu.dat};
    \addplot[draw=red_characteristic] table[x index=0, y index=16] {data/trajectory_0_15s_0_5_pu.dat};
    \addplot[draw=green_characteristic] table[x index=0, y index=16] {data/trajectory_0_25s_0_73_pu.dat};

    \addplot[draw=blue_characteristic, dashed] table[x index=0, y index=33] {data/trajectory_0_1s_0_2_pu.dat};
    \addplot[draw=red_characteristic, dashed] table[x index=0, y index=33] {data/trajectory_0_15s_0_5_pu.dat};
    \addplot[draw=green_characteristic, dashed] table[x index=0, y index=33] {data/trajectory_0_25s_0_73_pu.dat};
    \end{axis}
    
    \begin{axis}[
    xlabel near ticks, 
    ylabel near ticks, 
    width=\subplotWidth, 
    height=\subplotHeight,
    yshift=2*\heightOffset,
    xshift=0,
    line width=0.5,
    grid=major, 
    xlabel={Time $t$ [s]},
    xmin=0, xmax=1,
    axis x line=bottom,
    axis y line=left,
    ylabel={$Q_{total}$ [pu]},
    smooth,
    thick,
    ]

    \addplot[draw=blue_characteristic] table[x index=0, y index=17] {data/trajectory_0_1s_0_2_pu.dat};
    \addplot[draw=red_characteristic] table[x index=0, y index=17] {data/trajectory_0_15s_0_5_pu.dat};
    \addplot[draw=green_characteristic] table[x index=0, y index=17] {data/trajectory_0_25s_0_73_pu.dat};

    \addplot[draw=blue_characteristic, dashed] table[x index=0, y index=34] {data/trajectory_0_1s_0_2_pu.dat};
    \addplot[draw=red_characteristic, dashed] table[x index=0, y index=34] {data/trajectory_0_15s_0_5_pu.dat};
    \addplot[draw=green_characteristic, dashed] table[x index=0, y index=34] {data/trajectory_0_25s_0_73_pu.dat};
    
    \end{axis}

    \end{tikzpicture}
    \caption{System response to a voltage disturbance with characteristics $\mathcal{C} = [\Delta T, \Delta V]$. True response (dashed lines) and PINN approximation (solid). Two responses (blue and green) stem from the training dataset while the red one is previously unseen.}
    \label{fig:PINN_approximations}
\end{figure}

First, we consider the quality of the approximation of the PINN. \Cref{fig:PINN_approximations} presents the voltage disturbance $V_t$ and three critical variables, $ V_{\rm meas},  P_{\rm total},  Q_{\rm total}$ for three different disturbance characteristics $\mathcal{C}$. The blue and green predictions stem from trajectories that are included in the training dataset while the red curve is a previously unseen trajectory. All three trajectories are captured well, in particular in the initial phase until the disturbance is cleared, and for the later half when the system is mostly settled again. The approximation shows the largest errors around the clearing time which is due to the associated jumps in the algebraic variables that are more difficult to capture. In these regions, we can furthermore observe that the use of ReLUs as the activation function can lead to less smooth approximations due to the piece-wise linear characteristic of ReLUs. Additional neurons in the hidden layers can reduce these effects but they make the subsequent optimization computationally more expensive. 

\subsection{System analysis}
By solving the two optimization problems in \cref{subsec:optimization_problems} we obtain two different perspectives from which we can analyze the dynamic system. 
First, we can identify the boundary, i.e. the critical combinations of $\Delta V$ and $\Delta T$, for which we enter the LVRT control regime by considering when $V_{\rm meas}$ drops below $V_{int}=0.7$pu. 
The dashed lines show the ground truth based on a large number of time-domain simulations. While the approximation without additional conservativeness ($\epsilon = 0$pu) correctly identifies the boundary for $\Delta V < 0.4$, we observe a mismatch for larger disturbance magnitudes. This effect occurs because the PINN approximation overestimates the voltage nadir as it can also be seen for the green curve in \cref{fig:PINN_approximations}. By adding $\epsilon=0.025$pu or $\epsilon=0.05$pu we obtain more conservative approximations for the LVRT boundary.

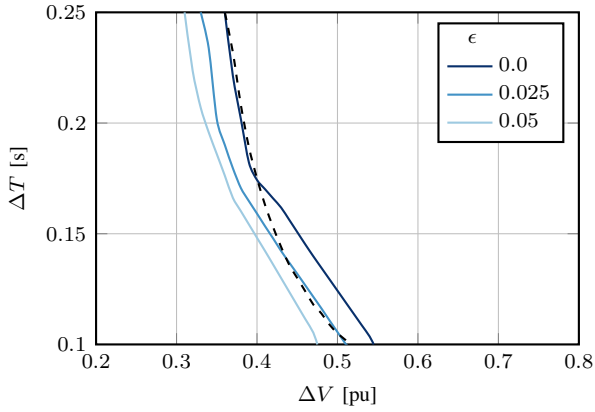
\begin{figure}[t]
    \centering
    \pagestyle{empty}
\def\subplotWidth{8cm}
\def\horizontalDistance{5cm}
\def\subplotHeight{6cm}
\def\heightOffset{-3.5cm}

\definecolor{epsilon_0}{HTML}{08306B}
\definecolor{epsilon_25}{HTML}{4292C6}
\definecolor{epsilon_50}{HTML}{9ECAE1}

    \begin{tikzpicture}[     every node/.style={font=\footnotesize}]

    \begin{axis}[
    xlabel near ticks, 
    ylabel near ticks, 
    width=\subplotWidth, 
    height=\subplotHeight,
    yshift=0,
    xshift=0,
    line width=0.5,
    grid=major, 
    xlabel={$\Delta V$ [pu]},
    ymax = 0.25,
    ymin= 0.1,
    xmin=0.2, xmax=0.8,
    ylabel={$\Delta T$ [s]},
    smooth,
    thick,
    legend pos=north east,
    legend cell align={left}]
   \addlegendimage{empty legend}
    \addplot[no marks, draw=epsilon_0] coordinates {(0.3600, 0.25000) 
    (0.3700, 0.22135) 
    (0.3800, 0.20172) 
    (0.3900, 0.18209) 
    (0.4000, 0.17429) 
    (0.4100, 0.17012) 
    (0.4200, 0.16596) 
    (0.4300, 0.16180) 
    (0.4400, 0.15632) 
    (0.4500, 0.15076) 
    (0.4600, 0.14520) 
    (0.4700, 0.13981) 
    (0.4800, 0.13463) 
    (0.4900, 0.12946) 
    (0.5000, 0.12429) 
    (0.5100, 0.11911) 
    (0.5200, 0.11394) 
    (0.5300, 0.10877) 
    (0.5400, 0.10359) 
    (0.545, 0.1)};
    
    \addplot[no marks, draw=epsilon_25]  coordinates {(0.33, 0.25) 
    (0.34, 0.2338) 
    (0.35, 0.2018) 
    (0.36, 0.1901) 
    (0.37, 0.1796) 
    (0.38, 0.1706) 
    (0.39, 0.1647) 
    (0.40, 0.1591) 
    (0.41, 0.1534) 
    (0.42, 0.1479) 
    (0.43, 0.1422) 
    (0.44, 0.1370) 
    (0.45, 0.1319) 
    (0.46, 0.1267) 
    (0.47, 0.1215) 
    (0.48, 0.1163) 
    (0.49, 0.1109) 
    (0.50, 0.1056) 
    (0.51, 0.1003) 
    (0.511, 0.1)};
    
    \addplot[no marks, draw=epsilon_50] coordinates { (0.3100, 0.25000) 
(0.3200, 0.22366) 
(0.3300, 0.20699) 
(0.3400, 0.19578) 
(0.3500, 0.18602) 
(0.3600, 0.17625) 
(0.3700, 0.16620) 
(0.3800, 0.16018) 
(0.3900, 0.15417) 
(0.4000, 0.14816) 
(0.4100, 0.14220) 
(0.4200, 0.13610) 
(0.4300, 0.12994) 
(0.4400, 0.12379) 
(0.4500, 0.11763) 
(0.4600, 0.11148) 
(0.4700, 0.10532) 
(0.475, 0.1)};

    \addplot[no marks, draw=black, dashed] coordinates {(0.36, 0.25) 
(0.37, 0.232) 
(0.38, 0.208) 
(0.39, 0.189) 
(0.4, 0.175) 
(0.41, 0.163) 
(0.42, 0.153) 
(0.43, 0.144) 
(0.44, 0.136) 
(0.45, 0.13) 
(0.46, 0.124) 
(0.47, 0.118) 
(0.48, 0.114) 
(0.49, 0.109) 
(0.5, 0.105) 
    (0.51, 0.102) 
    (0.52, 0.098)};

    \addlegendentry{\hspace{-.6cm}\textbf{$\epsilon$}}
    \addlegendentry{$0.0$}
    \addlegendentry{$0.025$}
    \addlegendentry{$0.05$}
    \end{axis}

    \end{tikzpicture}
    \caption{Predicted LVRT boundary by the optimization problem \eqref{eq:obj_maximumDT_1}-\eqref{eq:obj_maximumDT_2} based on the approximation of $\hat{V}_{\rm meas}$ and including the factor $\epsilon$ in pu to adjust the conservativeness. The dashed line represents ground truth.}
    \label{fig:StabilityBoundary}
\end{figure}

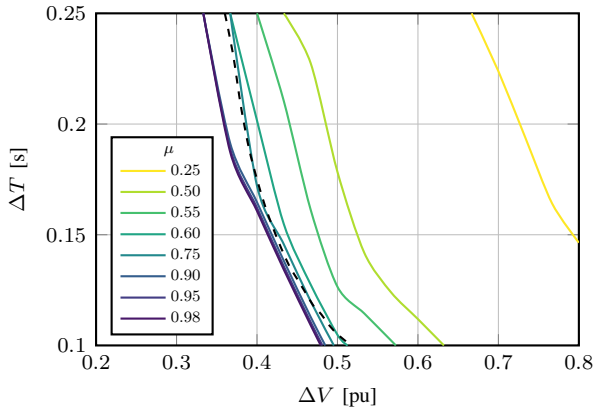
\begin{figure}[t]
    \centering
    \pagestyle{empty}
\def\subplotWidth{8cm}
\def\horizontalDistance{5cm}
\def\subplotHeight{6cm}
\def\heightOffset{-3.5cm}

    \begin{tikzpicture}[     every node/.style={font=\footnotesize}]

    \begin{axis}[
    xlabel near ticks, 
    ylabel near ticks, 
    width=\subplotWidth, 
    height=\subplotHeight,
    yshift=0,
    xshift=0,
    line width=0.5,
    grid=major, 
    xlabel={$\Delta V$ [pu]},
    ymax = 0.25,
    ymin= 0.1,
    xmin=0.2, xmax=0.8,
    ylabel={$\Delta T$ [s]},
    smooth,
    thick,
    colormap name=viridis,
    cycle list={[of colormap]},
    %
    legend pos=south west,
    legend style={nodes={scale=0.75, transform shape}}]
   \addlegendimage{empty legend}
    \addplot[no marks, index of colormap=17] coordinates { 
(0.6667, 0.25000)
(0.7000, 0.22384)
(0.7333, 0.19400)
(0.7667, 0.16415)
(0.8000, 0.14643)
};

    \addplot[no marks, index of colormap=15] coordinates { 
(0.4333, 0.25000)
(0.4667, 0.22748)
(0.5000, 0.17836)
(0.5333, 0.14186)
(0.5667, 0.12395)
(0.6000, 0.11190)
(0.6323, 0.1)
};

    \addplot[no marks, index of colormap=12] coordinates { 
(0.4000, 0.25000)
(0.4333, 0.21040)
(0.4667, 0.16128)
(0.5000, 0.12649)
(0.5333, 0.11445)
(0.5667, 0.10220)
(0.572504830454795, 0.1)
};

    \addplot[no marks, index of colormap=10] coordinates { 
(0.3667, 0.250000)
(0.4000, 0.201585)
(0.4333, 0.154835)
(0.4667, 0.127498)
(0.5000, 0.104791)
(0.512715469961012, 0.1)
};

    \addplot[no marks, index of colormap=8] coordinates { 
(0.3667, 0.25000)
(0.4000, 0.17207)
(0.4333, 0.14569)
(0.4667, 0.11995)
(0.495528681754353, 0.1)
};

    \addplot[no marks, index of colormap=5] coordinates { 
(0.3333, 0.25000)
(0.3667, 0.19126)
(0.4000, 0.16466)
(0.4333, 0.13823)
(0.4667, 0.11286)
(0.484910084600083, 0.1)
};

    \addplot[no marks, index of colormap=3] coordinates { 
(0.3333, 0.25000)
(0.3667, 0.18879)
(0.4000, 0.16220)
(0.4333, 0.13574)
(0.4667, 0.11053)
(0.481370552215324, 0.1)
};

    \addplot[no marks, index of colormap=1] coordinates { 
(0.3333, 0.25000)
(0.3667, 0.18731)
(0.4000, 0.16072)
(0.4333, 0.13425)
(0.4667, 0.10913)
(0.479290010774857, 0.1)
};

    \addplot[no marks, draw=black, dashed, thick] coordinates {(0.36, 0.25) 
(0.37, 0.232) 
(0.38, 0.208) 
(0.39, 0.189) 
(0.4, 0.175) 
(0.41, 0.163) 
(0.42, 0.153) 
(0.43, 0.144) 
(0.44, 0.136) 
(0.45, 0.13) 
(0.46, 0.124) 
(0.47, 0.118) 
(0.48, 0.114) 
(0.49, 0.109) 
(0.5, 0.105) 
    (0.51, 0.102) 
    (0.52, 0.098)};
    
    \addlegendentry{\hspace{-.6cm}\textbf{$\mu$}}
    \addlegendentry{$0.25$}
    \addlegendentry{$0.50$}
    \addlegendentry{$0.55$}    
    \addlegendentry{$0.60$}
    \addlegendentry{$0.75$}    
    \addlegendentry{$0.90$}
    \addlegendentry{$0.95$}    
    \addlegendentry{$0.98$}
    \end{axis}

    \end{tikzpicture}
    \caption{Power delivery $\hat{P}_{\rm total} = \mu P_{\rm ext}$ after 1s. The dashed line corresponds to the LVRT boundary and is associated with $\mu = 0.6$.}
    \label{fig:power_delivery}
\end{figure}

The second angle uses the fact that entering the LVRT control regime entails an immediate and in the short-term irreversible reduction of the reference currents. This, in turn, leads to a reduced delivered active power even after the disturbance is cleared. At the LVRT boundary we expect an immediate drop from a near full power delivery ($\mu = 1$) to about 60\% since $c = 0.6$ was chosen in \eqref{eq:partial_tripping}. \Cref{fig:power_delivery} shows that the power delivery is very sensitive to small changes in the disturbance around the LVRT boundary. 

\section{Conclusion} \label{sec:conclusion}
This work presents a rigorous and flexible framework that can deliver insights on critical system responses, and determine critical indices such as the critical clearing time after a power system disturbance, while avoiding the need for exhaustive time-domain simulations. We use Physics Informed Neural Networks (PINN) to accurately capture the underlying power system dynamics, and through an exact transformation, we reformulate the PINN to a mixed integer linear program. This provides for a tractable optimization program that captures in its constraints the power system dynamics, initially encoded in the PINN. The presented results show how we can determine the critical clearing time, or the delivered active power at the Low-Voltage-Ride-Through boundary without the need for exhaustive time domain simulations. Our methods also account for the potential approximation error of the PINNs by introducing an $\epsilon$-conservativeness factor, which can also be applied to account for a required stability margin. Future work will look into neural network verification methods that can drive neural network training towards the minimum PINN approximation error.

\bibliographystyle{myIEEEtran.bst}
\bibliography{References.bib}{}

\appendix
\begin{align}
    v_{d} &= v_{x}\cos(\theta_{{\rm pll}}) + v_{y}\sin(\theta_{{\rm pll}}), \label{Eq:vd_vx_vy}\\
    v_{q} &= -v_{x}\sin(\theta_{{\rm pll}}) + v_{y}\cos(\theta_{{\rm pll}}), \label{Eq:vq_vx_vy}\\
    \frac{d \theta_{{\rm pll}}}{dt} &=(\omega_{{\rm pll}})\omega_{\rm ref},\label{Eq:theta} \\
    \omega_{{\rm pll}} &= K_{{p\omega}}v_{q} +  1,\\
    P_{\rm VSC} &= v_{d}i_{d} + v_{q}i_{q}, \\
    Q_{\rm VSC} &= v_{q}i_{d} - v_{d}i_{q}.\\
     v_{d} &=  v_{g,d} - \omega_{{\rm pll}}L_c i_{q} + R_c i_{d},  \\
    v_{q} &=  v_{g,q} + \omega_{{\rm pll}}L_c i_{d} + R_c i_{q}
\end{align}

\end{document}